\documentclass[trackchanges, twocolumn]{aastex7}
\usepackage{amsmath}

\begin{document}

\title{A single power law for the TRAPPIST-1 flare distribution across four orders of magnitude in energy} 

\author[0009-0009-3020-3435]{Valeriy Vasilyev}
\affiliation{Max Planck Institute for Solar System Research, Justus-von-Liebig-Weg 3, 37077 G\"ottingen, Germany
}
\email[show]{vasilyev@mps.mpg.de }  
\author[0000-0002-8842-5403]{Alexander I. Shapiro}
\affiliation{Institute of Physics, University of Graz, 8010 Graz, Austria
}
\affiliation{Max Planck Institute for Solar System Research, Justus-von-Liebig-Weg 3, 37077 G\"ottingen, Germany
}
\email[]{}
\author[0000-0002-6087-3271]{Nadiia Kostogryz}
\affiliation{Max Planck Institute for Solar System Research, Justus-von-Liebig-Weg 3, 37077 G\"ottingen, Germany
}
\email[]{}  
\author[0000-0001-5989-7594]{Chia-Lung Lin}
\affiliation{Department of Astronomy Steward Observatory, The University of Arizona, 933 N. Cherry Avenue, Tucson, AZ 85721, USA}
\email[]{}  
\author[]{Greg Kopp}
\affiliation{Laboratory for Atmospheric and Space Physics, University of Colorado Boulder, 3665 Discovery Drive, Boulder, CO 80303, USA;}
\affiliation{Institute of Physics, University of Graz, 8010 Graz, Austria
}
\email[]{}
\author[0000-0002-3627-1676]{Benjamin V. Rackham}
\affiliation{Department of Earth, Atmospheric and Planetary Sciences, Massachusetts Institute of Technology, Cambridge, MA 02139, USA}
\affiliation{Kavli Institute for Astrophysics and Space Research, Massachusetts Institute of Technology, Cambridge, MA 02139, USA}
\email[]{}
\author[0000-0003-2073-002X]{Astrid M. Veronig}
\affiliation{Institute of Physics, University of Graz, 8010 Graz, Austria
}
\affiliation{University of Graz, Kanzelh\"ohe Observatory for solar and Environmental Research, Kanzelh\"ohe 19, 9521 Treffen, Austria
}
\email[]{}

\author[0000-0003-4676-0622]{Olivia Lim}
\affiliation{Institut Trottier de recherche sur les exoplan\`etes, Université de Montréal, 1375 Ave Thérèse-Lavoie-Roux, Montréal, QC, H2V 0B3, Canada}
\email[]{}

\author[0000-0003-2415-2191]{Julien de Wit}
\email{jdewit@mit.edu}
\affiliation{Department of Earth, Atmospheric and Planetary Sciences, Massachusetts Institute of Technology, 77 Massachusetts Avenue, Cambridge, MA 02139, USA}

\author[0000-0003-3714-5855]{Daniel Apai}
\email{apai@arizona.edu}
\affiliation{University of Arizona, 933 N.\ Cherry Avenue, Tucson, AZ 85721, USA}

\author[0000-0001-7696-8665]{Laurent Gizon}
\affiliation{Max Planck Institute for Solar System Research, Justus-von-Liebig-Weg 3, 37077 G\"ottingen, Germany
}
\affiliation{Institut für Astrophysik und Geophysik, Georg-August-Universit\"at G\"ottingen,  37077  G\"ottingen, Germany
}
\email[]{}
\author[0000-0002-3418-8449]{Sami K. Solanki}
\affiliation{Max Planck Institute for Solar System Research, Justus-von-Liebig-Weg 3, 37077 G\"ottingen, Germany
}
\affiliation{School of Space Research, Kyung Hee University, Yongin, Gyeonggi 17104, Republic of Korea}
\email[]{}
\author[0000-0002-6892-6948]{Sara Seager}
\affiliation{Department of Physics,
Massachusetts Institute of Technology, Cambridge, MA 02139, USA}
\affiliation{Department of Earth, Atmospheric and Planetary Sciences, Massachusetts Institute of Technology, Cambridge, MA 02139, USA}
\affiliation{Kavli Institute for Astrophysics and Space Research, Massachusetts Institute of Technology, Cambridge, MA 02139, USA}
\affiliation{Department of Aeronautics and Astronautics,
MIT, 77 Massachusetts Avenue, Cambridge, MA 02139, USA}
\email[]{}

\begin{abstract}
TRAPPIST-1 is an ultra-cool dwarf that flares frequently. These flares shape the surrounding planets' high-energy irradiation environments, with consequences for atmospheric chemistry and escape, and they can contaminate transmission spectroscopy of those planets. A quantitative flare-frequency distribution (FFD) spanning the full energy range is therefore essential for both interpreting JWST spectra and modeling the planets' irradiation histories. Here we present a unified FFD over four orders of magnitude in energy by jointly analyzing $\approx$87\,hr of JWST/NIRISS and JWST/NIRSpec time-series spectroscopy together with $\approx$74\,days of   \textit{Kepler}/K2 photometry. To enable a consistent comparison across these heterogeneous datasets, we convert all events to energies in the TESS bandpass. For the Kepler-to-TESS conversion we adopt a cooler flare continuum appropriate for ultra-cool dwarfs ($T_{\rm flare}=3500$\,K). After correcting for flare-detection sensitivities, the combined JWST+K2 cumulative FFD is consistent with a single power law, $N(\ge E_\mathrm{TESS})\propto E_\mathrm{TESS}^{-\beta}$, with $\beta=0.753$ over $E_{\rm TESS}\simeq10^{29}$--$10^{33}$\,erg. 
The slope of the distribution indicates that the time-averaged flare energy budget is dominated by rare, high-energy events rather than by the more numerous low-energy flares. 
Moreover, we found that strong flares with energies $E_\mathrm{TESS} > 10^{32}$~erg occur once every 25 days, about an order of magnitude more frequently than inferred from previous TRAPPIST-1/analog FFD estimates. This elevated rate of energetic flares has important implications for atmospheric escape, photochemistry, and habitability assessments of the TRAPPIST-1 planets.
\end{abstract}


\section{Introduction}
TRAPPIST-1 is a nearby ultra-cool M-type dwarf star that hosts a compact planetary system. It is one of the key targets for James Webb Space Telescope (JWST) transit spectroscopy because three of its seven known planets lie within the habitable zone \citep{Gillon2016, Gillon2017}. Orbiting a magnetically active star \citep{Luger2017, Morris2018}, TRAPPIST-1 planets are exposed to intense radiation from stellar flares and potentially to an energetic particle environment \citep{Garraffo2017, Roettenbacher2017, Vida2017, Paudel2018, Howard2023, Howard2025}. Furthermore,  subtle atmospheric signatures of TRAPPIST-1 planets are overwhelmed by strong activity signals. While this  hampers  atmospheric characterization \citep{Rackham2018, Zhang2018, Wakeford2019, Garcia2022, Lim2023, Espinoza2025, Glidden2025, Piaulet-Ghorayeb2025, Radica2025, Rathcke2025, Allen2025}, it also makes TRAPPIST-1 an ideal laboratory for studying magnetic activity in cool  M dwarfs and star–planet interactions \citep{Gillon2025}. In particular, a substantial effort has been recently invested in studying its flaring activity \citep{Howard2023, Howard2025, Vasilyev2025}.  

Cool red dwarfs have been known to be notoriously flaring active since the advent of large space-based photometric surveys. In particular, the Transiting Exoplanet Survey Satellite (TESS) \citep{Ricker2015} and \textit{Kepler} \citep{Borucki2010} have provided massive amounts of data to understand the flare statistics on M dwarfs \citep{Hawley2014, Davenport2016, Yang2017, Guenter2020, Feinstein2020}.  They have revealed that  the cumulative flare frequency distribution (FFD) on M dwarfs can be described by a power law \citep{Aschwanden2021, Feinstein2022}, i.e. $\nu(>E)\sim E^{-\beta}$, where $\beta$ is the slope of the cumulative distribution.   

Direct photometric observations of TRAPPIST-1 across multiple studies also reveal power-law flare-frequency distributions. For example, analyzing $\sim$ 74 days of \textit{Kepler} K2 observations of the system, \cite{Vida2017} identified 42 flares with energies ranging from $1.26\times10^{30}$ to $1.24\times10^{33}$~erg and derived a power-law slope of $\beta=\alpha-1 \approx 0.59$ ($\alpha$ is the slope of the differential FFD). Using \textit{Kepler} K2 data, \cite{Paudel2018} analyzed 39 flares spanning $6.5\times10^{29}$ to $7.2\times10^{32}$ erg and found a slope of $\beta=0.6$.  \citet{Ducrot2020} detected five flares in \textit{Spitzer} observations at 3.6 and 4.5~$\mu$m and measured a similar slope.   Later, \cite{Seli2021} extended analysis on stars similar to TRAPPIST-1 using photometric time series from TESS, obtaining the power-law index of $\beta\sim1$.   

Recent high-cadence spectroscopic observations reveal more frequent flaring than previously detected. \cite{Howard2023} reported four flares during JWST observations, measuring a flaring rate an order of magnitude  higher than previous photometric observations. A further extension of this analysis to six flares yielded a cumulative FFD slope of $\beta=1.26$ \citep{Howard2025}.
Taken together, these results indicate that flare rates and FFD slopes reported for TRAPPIST-1 are not fully consistent across different datasets and methodologies.

A significant source of uncertainty in estimating FFD can be attributed to poorly constrained  flare temperatures. 
This is because flare temperature defines the relationship between energies measured in broad-band photometric passband (e.g., \textit{Kepler} and TESS) and bolometric energy. In the absence of direct measurements of flare temperatures on TRAPPIST-1 before the JWST era, flare temperatures of ultra-cool dwarfs have typically been assumed to match those of solar white-light flares, being around 9000–10,000 K \citep[][]{Kretzschmar2011}.  However, JWST spectroscopic observations showed that the effective temperature of flares in TRAPPIST-1 is around 3500--5300\,K \citep{Howard2023, Howard2025}. Also, ground-based multicolor photometric observations revealed that flare temperatures on M-dwarfs are significantly lower \citep{Maas2022, Lin2025} than on G-type stars.  From a physical point of view, in the cool, dense atmosphere of TRAPPIST-1, magnetic heating is strongly moderated by the dissociation of molecular hydrogen (H$_2$), which acts as an efficient energy sink, preventing white-light flare regions from heating above ${\sim}3500$\,K \citep{ThermostatShapiro}.

In this letter, we perform a self-consistent analysis of TRAPPIST-1 flares observed by the JWST (NIRISS and NIRSpec) and \textit{Kepler} K2 missions.  In Section~\ref{sec:jwst} and \ref{sec:k2}, we describe how flare energies have been measured in these datasets by  adopting a $T_{\rm flare}=3500$ K flare temperature for K2 data and converting K2 bandpass energies into the TESS band. In Section~\ref{sec:ffd}, we show the measured FFD, and Section~\ref{sec:discussion} summarizes and provides a discussion of our results.

\section{Flare energies}
Our goal is to measure TRAPPIST-1 flare energies in the JWST(NIRISS/NIRSpec) and \textit{Kepler}/K2 datasets and place both on a common scale for consistent comparisson by converting to the TESS bandpass, adopted here as a common reference band, to enable a consistent comparison across datasets. We then construct a cumulative FFD in that band.

\subsection{JWST observations}\label{sec:jwst}
We searched for stellar flares in calibrated JWST time-series spectroscopic observations of TRAPPIST-1, spanning a total observing time of $\Delta T^{\mathrm{(JWST)}}_{\mathrm{obs}} \approx 87$\,hr obtained during transit observations. A detailed summary of the analyzed JWST observations and data products is provided in Appendix~\ref{app:data}  Table~\ref{tab:obs_jwst_summary}.

For JWST/NIRISS/SOSS, we analyzed reduced time-series spectra provided by Lim et al. (in prep.; private communication), generated using the reduction procedures described in \citet{Lim2023}.
For JWST/NIRSpec PRISM, we analyzed time-series observations reduced and calibrated by Lin et al. (in prep.) using a uniform procedure across all visits; the main reduction and calibration steps are summarized in Appendix~\ref{app:b}.

A subset of these data has previously been discussed in the context of stellar flares by \citet{Howard2023, Vasilyev2025, Howard2025}, and is further analyzed in Lin et al. (in prep.).
Of the total observing time, $\approx35$\,hr were obtained with NIRISS and $\approx52$\,hr with NIRSpec/PRISM.
The analyzed JWST/NIRISS/SOSS data were provided at a cadence of $\approx 104$\,s; for consistency, we rebinned the NIRSpec/PRISM data to the same cadence.

\begin{figure*}[htb]
  \centering
  \includegraphics[width=\textwidth]{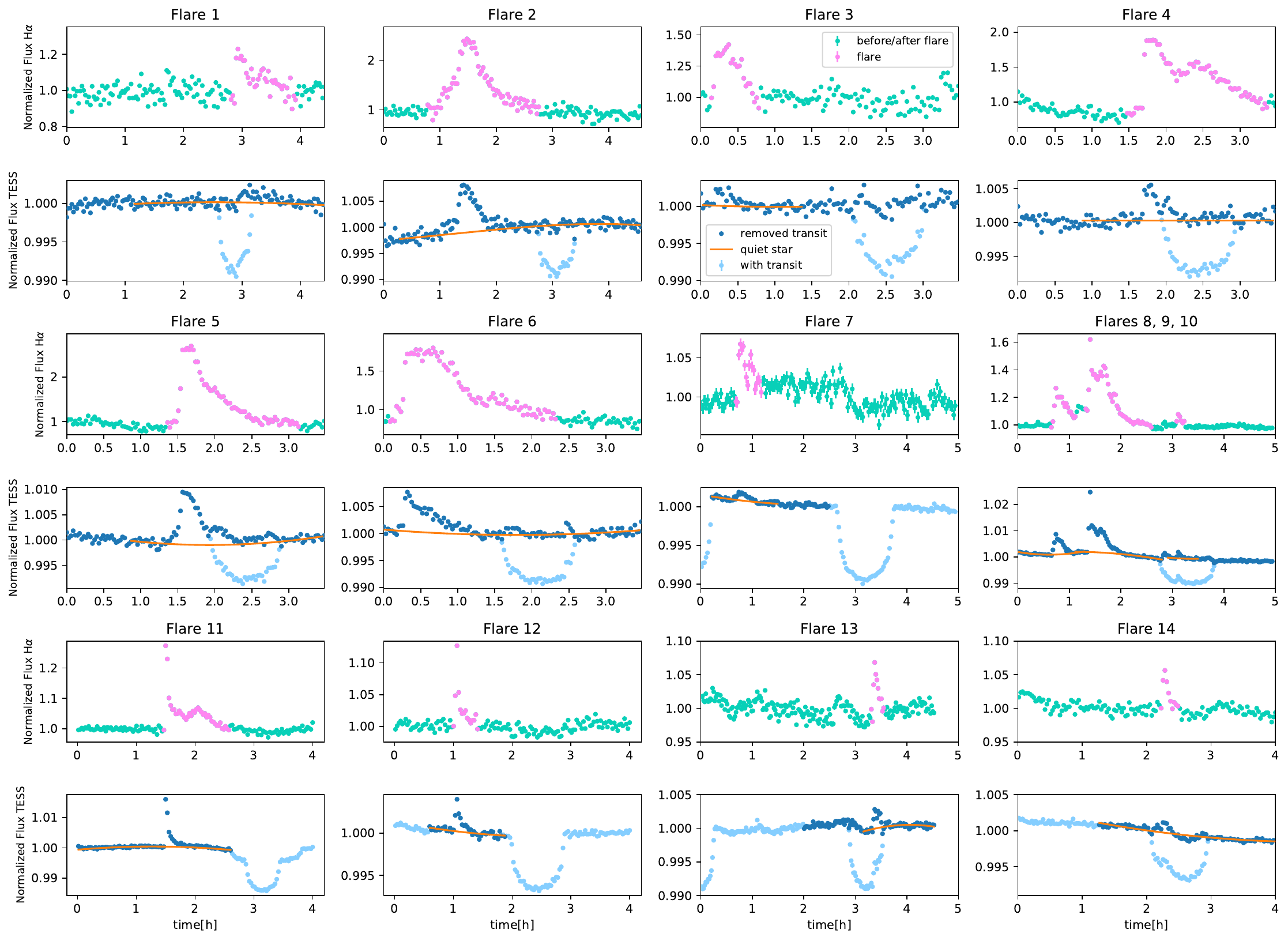}
\caption{\textbf{Flares observed with JWST.}
Rows 1, 3, and 5 show the normalized H$\alpha$ line-flux light curves used to define the flare windows (teal: points outside the flare window; magenta: points inside the flare window).
Rows 2, 4, and 6 show the corresponding JWST-derived TESS-band light curves for the same events, including the data before transit removal (light blue), after transit removal (dark blue), and the low-order polynomial baseline adopted for the quiescent stellar flux during the flare (orange). Flares~1--6 were observed with NIRISS and Flares~7--14 with NIRSpec. All light curves are normalized to their median flux at non-flaring times.}
  \label{fig:jwst_niriss}
\end{figure*}

From the spectral time series, we extracted light curves centered on H$\alpha$ and identified 14 flares (hereafter Flares 1--14). Flare start and end times were determined when the H$\alpha$ flux leaves and when it  returns to the pre-flare level, respectively. Because our reference band is TESS, we projected the JWST spectra into the TESS bandpass by weighting each spectrum with the TESS transmission function  $R_\mathrm{TESS}(\lambda)$ at each time. Hereafter, $F^{(\mathrm{JWST})} (t)$ denotes the light curve in the TESS bandpass.  Light curves in H$\alpha$ and TESS bandpasss are shown in Figure ~\ref{fig:jwst_niriss}.  Since the primary goals of these observations were transits, in some observations a flare overlaps in time with a transit. For example, the maximum light phases of Flares 1, 10, 13, and 14 were during planetary transits, and for several flares (4, 5, and 6) the flare decay phase occurred during the transit. 

We removed planetary transits from the light curves where flares overlapped the transits. To do so, we modeled a transit using the \texttt{batman} package \citep{Kreidberg2015}. For flare decay phases overlapping transits, we modeled the local stellar baseline with an exponential tail, representing the flare decay phase.  
For the dataset in which the flare peak occurred during transit, the transit depth could not be constrained robustly when fitted simultaneously with the flare profile. We therefore adopted the best-fit transit model (and corresponding planetary radius) from a separate, flare-free observation of the same planet. This model was evaluated at the timestamps of the flare-contaminated dataset and removed to produce a transit-free light curve for subsequent flare characterization.

Using data segments bracketing each flare event, we fit a low-order polynomial (1 or 2) to estimate the quiescent stellar flux integrated in the TESS band $F^{(\mathrm{JWST})}_{0}(t)$ (see orange curves in Fig.~\ref{fig:jwst_niriss}).  We then define the equivalent duration,
\begin{equation}
\mathrm{ED}^{(\mathrm{JWST})}_\mathrm{TESS}
= \int_{t_{\rm start}}^{t_{\rm end}} \frac{\Delta F^{(\mathrm{JWST})}(t)}{F^{(\mathrm{JWST})}_{0}(t)}\,\mathrm{d}t,
\end{equation}
where  $\Delta F^{(\mathrm{JWST})}(t) = F^{(\mathrm{JWST})}(t)-F^{(\mathrm{JWST})}_{0}(t)$ is the flux excess from the flare.  $\mathrm{ED}$ has units of time and measures the flare fluence in the band relative to the quiescent stellar flux.

The corresponding TESS-band flare energy is
\begin{equation}
E^{(\mathrm{JWST})}_{\mathrm{TESS}} \;=\; 4\pi d^{\,2}\,\langle F^\star_{\mathrm{TESS}}\rangle \,\mathrm{ED}^{(\mathrm{JWST})}_\mathrm{TESS},
\end{equation}
where $d=12.467\pm0.011$\,pc is the Gaia~DR3 distance \citep{GaiaCollaboration2023} and $\langle F^\star_{\mathrm{TESS}}\rangle$ is the time-averaged quiescent TESS-band flux, computed by averaging the JWST-derived TESS-band flux outside of transits and flares.
Uncertainties in $E^{(\mathrm{JWST})}_{\mathrm{TESS}}$ include contributions from the photometric scatter about the fitted baseline and the distance uncertainty. 

\subsection{\textit{Kepler}/K2 observations}\label{sec:k2}
We analyzed the 1\,min-cadence pre-search data conditioning simple aperture photometry  (PDCSAP) \textit{Kepler}/K2 light curve with total duration 
$\Delta T^{\mathrm{(K2)}}_{\mathrm{obs}}=73.65$\,d.
To identify flare candidates, we removed long-timescale variability by iteratively applying a running-mean (boxcar) filter with a 0.04\,d window, adopted to be long relative to flare durations but short enough to remove slow variability.  We then subtracted the smoothed trend from the original light curve and performed iterative $3\sigma$ clipping to mask outliers. The clipping procedure was repeated for 10 iterations, after which the set of masked cadences had converged. 
We estimated the noise $\sigma$ as the standard deviation of the detrended, outlier-masked light curve.
Flare candidates were flagged as sequences with $\geq2$ consecutive cadences exceeding $3\sigma$ above the local baseline and then visually vetted.
This yielded 44 flares, including several complex, multi-peaked events.

For each flare we extracted a ${\sim}0.1$\,d window centered on the event and performed a \emph{simultaneous} fit of the quiescent baseline and flare profile.
The quiescent signal $C_0(t)$ was modeled as a first-order polynomial.
Flare morphology was modeled using the empirical \textit{Kepler}-band template of \citet{Davenport2014}; multi-peaked events were modeled as a sum of template components.
Parameter inference was performed using an Markov Chain Monte Carlo sampler \textsc{EMCEE} \citep{emcee2013}, and posterior samples were propagated to derived quantities.

We then compute the equivalent duration from the fitted fractional excess in the \textit{Kepler} light curve,
\begin{equation}
\mathrm{ED}^{(\mathrm{K2})}_{K}
= \int_{t_{\rm start}}^{t_{\rm end}} \frac{\Delta C(t)}{C_0(t)}\,\mathrm{d}t,
\qquad
\Delta C(t)=C(t)-C_0(t),
\end{equation}
where $C(t)$ denotes the PDCSAP signal (proportional to detected photoelectron count rate) and $\mathrm{ED}^{(\mathrm{K2})}_{K}$ again has units of time.

Because \textit{Kepler} is a photon-counting instrument, the band-integrated signal is naturally photon-weighted.
To relate the \textit{Kepler} fractional amplitude (measured from the count-rate light curve) to an energy-flux convention, we write the flare excess spectrum observed at Earth as $F_\lambda^{\rm flare}(\lambda,t)=A(t)\,f_\lambda(\lambda)$, where $f_\lambda(\lambda)$ is the flare spectral shape and $A(t)$ is a scalar normalization that captures the time-variable strength of the flare.
Hereafter $F_\lambda^\star(\lambda)$ denotes the quiescent stellar spectral flux density at Earth (i.e., the spectrum in the absence of flares), in the same units as $F_\lambda^{\rm flare}$.

With this notation,
\begin{equation}
\left(\frac{\Delta C}{C_0}\right)_{K}(t)
=
A(t)\,
\frac{\int f_\lambda(\lambda)\,\lambda\,R_{K}(\lambda)\,\mathrm{d}\lambda}
     {\int F_\lambda^\star(\lambda)\,\lambda\,R_{K}(\lambda)\,\mathrm{d}\lambda},
\end{equation}
where $R_{K}(\lambda)$ is the \textit{Kepler} filter response and the factor of $\lambda$ reflects photon counting (the constant $(hc)^{-1}$ cancels in the ratio).
If instead one defines the band-integrated fractional amplitude in \emph{energy} units,
\begin{equation}
\left(\frac{\Delta F}{F_0}\right)_{K}(t)
=
A(t)\,
\frac{\int f_\lambda(\lambda)\,R_{K}(\lambda)\,\mathrm{d}\lambda}
     {\int F_\lambda^\star(\lambda)\,R_{K}(\lambda)\,\mathrm{d}\lambda},
\end{equation}

Then the two conventions are related by
\begin{equation}
\epsilon =  \frac{(\Delta F/F_0)_{K}}{(\Delta C/C_0)_{K}}
= \frac{\int f_\lambda R_{K}\,\mathrm{d}\lambda}{\int f_\lambda \lambda R_{K}\,\mathrm{d}\lambda}
  \times
  \frac{\int F_\lambda^\star \lambda R_{K}\,\mathrm{d}\lambda}{\int F_\lambda^\star R_{K}\,\mathrm{d}\lambda},
\label{eq:epsK}
\end{equation}
which is constant in time if the flare spectral shape does not change in time. We adopt $f_\lambda$ as a blackbody at $T_{\rm flare}\approx 3500$~K,  hereafter $B_\lambda(\lambda,T_{\rm flare})$. This flare temperature choice reflects the "thermostat" effect reported for ultracool-dwarf flares in the near-IR: a substantial fraction of the deposited flare energy may be diverted into molecular dissociation rather than further heating, limiting increases in the effective flare continuum temperature \citep{ThermostatShapiro}. This behavior has also been observed in JWST time-series spectroscopy \citep{Howard2023,Howard2025}. 

As a quiet stellar spectrum $F_\lambda^\star$ we use the Gaia~DR3 XP observed spectrum for the TRAPPIST-1 star that covers the \textit{Kepler} filter wavelength range. Thus the \emph{energy-based} equivalent duration is
\begin{equation}
\mathrm{ED}^{(K2, Energy)}_{K} \;=\; \epsilon\, \mathrm{ED}^{(K2)}_{K}.
\end{equation}

Given $R_{K}(\lambda)$, the band-integrated quiescent flux and luminosity are
\begin{equation}
F^{\star}_{K} = \int F_\lambda^\star(\lambda)\, R_{K}(\lambda)\,\mathrm{d}\lambda,
\qquad
L^{\star}_{K} = 4\pi d^{\,2} F^{\star}_{K},
\end{equation}
and the \textit{Kepler}-band flare energy is
\begin{equation}
E^{(K2)}_K \;=\; \mathrm{ED}^{(K2,\mathrm{Energy})}_{K}\, L^{\star}_{K}.
\label{eq:EK}
\end{equation}

We report flare energies in the TESS band as a common reference.
Using the notation introduced above, $F_\lambda^{\rm flare}(\lambda,t)=A(t)\,f_\lambda(\lambda)$, the band-integrated flare excess flux in bandpass $X$ is
\begin{equation}
F^{\rm flare}_{X}(t) \;=\; \int F_\lambda^{\rm flare}(\lambda,t)\,R_X(\lambda)\,\mathrm{d}\lambda, 
\qquad X\in\{K,TESS\}.
\end{equation}
If the flare spectral shape is time-invariant and approximated as a blackbody, then 
$f_\lambda(\lambda)=B_\lambda(\lambda,T_{\rm flare})$. This approximation is commonly adopted for optical/near-IR flare analyses because broad-band flare emission is often dominated by a hot "white-light" continuum, while typical photometric data do not provide sufficient spectral information to robustly constrain a time-dependent flare temperature $T_{\rm flare}(t)$ or its dependence on flare properties (e.g., amplitude, duration, and energy). Recent work has demonstrated that time-dependent effective flare temperatures can be inferred when time-resolved spectroscopy is available and the flare continuum and line emission are modeled with radiative-hydrodynamic simulations \citep{Howard2025}. Accordingly, in this work we adopt a time-invariant spectral shape parameterized by a single effective temperature, then the ratio of band-integrated flare excess fluxes is constant in time and the corresponding flare energies satisfy
\begin{equation}
E_\mathrm{TESS}^{(K2)} \;=\; E_{K}^{(K2)} \,
\frac{\int B_\lambda(\lambda, T_{\rm flare})\, R_\mathrm{TESS}(\lambda)\,\mathrm{d}\lambda}
     {\int B_\lambda(\lambda, T_{\rm flare})\, R_{K}(\lambda)\,\mathrm{d}\lambda}.
\label{eq:ET_from_EK}
\end{equation}

To estimate uncertainties in $E_{T}^{(K2)}$, we propagate the uncertainty in $E_{K}^{(K2)}$, dominated by photometric noise and continuum placement in $\mathrm{ED}^{(K2)}_{K}$, the adopted quiescent spectrum $F_\lambda^\star$, and the distance $d$.

\subsection{Flare–frequency distribution}\label{sec:ffd}
\begin{figure*}[htbp]
  \centering
  \includegraphics[width=0.9\linewidth]{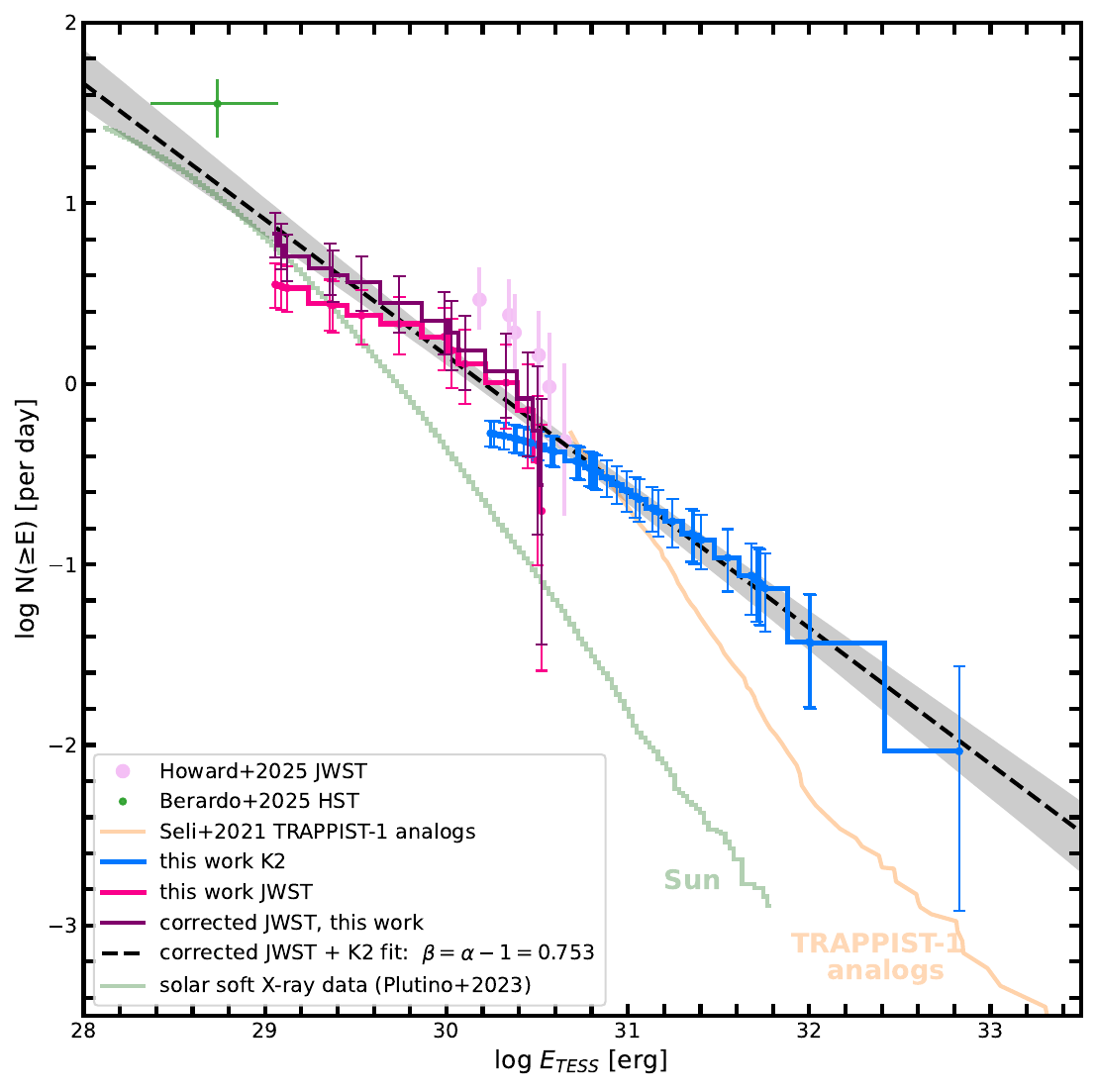}
  \caption{\textbf{Cumulative flare-frequency distribution of TRAPPIST-1.}
  The cumulative occurrence rate $\nu(\ge E_\mathrm{TESS})$ is shown as a function of TESS-band flare energy.
  The K2 FFD from this work is shown in blue, the observed JWST FFD in magenta, and the JWST FFD after applying the injection--recovery completeness correction in purple. The light green curve shows the solar soft X-ray flare distribution, derived from the catalog of \cite{Plutino2023} and converted to the TESS band assuming $T_\mathrm{flare}=9000$~K \citep{Kretzschmar2011}. 
The corrected JWST and K2 samples are consistent with a single power law in $E_{\rm TESS}$ over $\sim$4 orders of magnitude (dashed black line), with the gray band indicating the $1\sigma$ bootstrap uncertainty. At flare energies above $\sim 10^{29}$ erg, the flare occurrence rate of TRAPPIST-1 exceeds that of the Sun.}
  \label{fig:ffd}
\end{figure*}

In Figure~\ref{fig:ffd} for both datasets (JWST and K2) we show the measured cumulative FFD.  For each dataset, we sort events by energy, form the empirical cumulative counts $n(\ge E_\mathrm{TESS})$, and set the flaring rate $\nu=n/\Delta T_{\mathrm{obs}} $ with $\Delta T_{\mathrm{obs}}$ the duration of observations ($\Delta T^{\mathrm{(K2)}}_{\mathrm{obs}}$ for K2 and $\Delta T^{\mathrm{(JWST)}}_{\mathrm{obs}}$ for JWST ). Uncertainties in $\nu$ include Poisson counting errors; for K2, we also propagate energy uncertainties by Monte Carlo, perturbing $E_\mathrm{TESS}$ by its errors and adding the resulting scatter in quadrature.  For the 14 flares detected in JWST data, due to the low number of events, we include Poisson counting errors. 

For the limited number of detected flares in the JWST dataset, we estimate the energy-dependent detection completeness via injection--recovery experiments. Specifically, we inject synthetic flares of a given energy into the H$\alpha$ light curves and apply a flare detection pipeline to measure the completeness curve $p(E_\mathrm{TESS})=N_{\mathrm{recovered}}(E_\mathrm{TESS})/N_{\mathrm{injected}}(E_\mathrm{TESS})$ (see details in Appendix \ref{sec:appendix_detection_eff_jwst}). Using this completeness curve, we compute the completeness-corrected cumulative FFD as
\begin{equation}
\nu_{\mathrm{corr}}^{(\mathrm{JWST})}(>E_\mathrm{TESS})=\frac{n_{\mathrm{true}}(>E_\mathrm{TESS})}{\Delta T^{(\mathrm{JWST})}_{\mathrm{obs}}},
\end{equation}
where 
\begin{equation}
n_{\mathrm{true}}(>E_\mathrm{TESS})=\sum_{i:\,E_i\ge E_\mathrm{TESS}}\frac{1}{p(E_i)}
\end{equation}
is an estimator of the expected true number of flares above a threshold, accounting for energy-dependent incompleteness.

For the K2 dataset, the sample is effectively complete at high energies. We therefore restrict the analysis to the linear high-energy tail of the cumulative FFD, selecting flares with $E^{(\mathrm{K2})}_\mathrm{TESS}\ge E_{\min}$, where $E_{\min}=3.6\times 10^{30}$~erg (see Appendix \ref{app:emin} for the choice of $E_{\min}$), and compute the cumulative flare rate as 
$\nu^{(\mathrm{K2})}(>E_\mathrm{TESS})=N(>E_\mathrm{TESS})/\Delta T^{(\mathrm{K2})}_{\mathrm{obs}}$.

We model the cumulative FFD of the combined JWST and K2 datasets with a single power law,
\begin{equation}
\nu(E>E_\mathrm{TESS}) = A\left(\frac{E}{E_0}\right)^{-\beta},
\end{equation}
where $\beta$ is the cumulative power-law slope and $A$ is the normalization defined as the cumulative flare rate at the pivot energy $E_0$, i.e.\ $A = \nu(E>E_0)$. We adopt $E_0 = 1.01\times10^{31}\,\mathrm{erg}$ (the median energy of the combined flare sample) to reduce the covariance between the slope and normalization.

Because the two datasets contain different numbers of flares (and hence different numbers of cumulative points), we use a dataset-balanced weighting scheme in which each dataset contributes equal total weight to the fit. Concretely, we assign 50\% of the total weight to the K2 cumulative points and 50\% to the JWST cumulative points, distributing that weight uniformly within each dataset, so that neither dataset dominates the result solely because it contains more points. For fixed $E_0$ we obtain $\beta = 0.753^{+0.068}_{-0.050}$ and $A=\nu(>E_0)=0.240^{+0.035}_{-0.035}\,\mathrm{day}^{-1}$, where uncertainties correspond to the 16th--84th percentile range. Uncertainties on $\beta$ and $A$ are estimated via bootstrap resampling of the underlying flare lists: in each realization we resample the K2 flares (above $E_{\min}$) and the JWST flares with replacement (preserving the associated $p(E_\mathrm{TESS})$ values), recompute the cumulative FFD points, refit the power law with the same 50/50 dataset weighting, and derive confidence intervals from the resulting parameter distributions.

\section{Discussion and Conclusions}\label{sec:discussion}
\subsection{A Unified Flare-Frequency Distribution for TRAPPIST-1}
Figure~\ref{fig:ffd} summarizes our main result: the JWST and K2 cumulative FFDs are consistent with a \emph{single} power law across the energy range probed by the combined data. The best-fit cumulative slope is $\beta = 0.753$ (corresponding to a differential index $\alpha=\beta+1=1.753$ for $d\nu/dE\propto E_\mathrm{TESS}^{-\alpha}$), shown by the dashed line with the shaded bootstrap uncertainty band.  Because $\alpha<2$, the time-averaged flare energy output is dominated by the rarest, highest-energy events. This behavior contrasts with the Sun, where $\alpha>2$ (with $\alpha\approx2.4$; \citealt{Vasilyev2024}), implying that the energy output is dominated by frequent, low-energy flares. Although the connection between white-light flares and particle events on TRAPPIST-1 remains poorly constrained, our result suggests that, unlike the Sun, rare flares may dominate particle-associated perturbations.

Power-law flare-frequency distributions are commonly discussed in the framework of self-organized criticality (SOC), in which magnetic systems release stored magnetic energy through avalanche-like cascades over a broad range of scales. The avalanche model explains the power-law distribution of solar flares \citep{Lu1993}. Recent high-resolution Solar Orbiter observations have provided direct observational support for avalanche-like reconnection during a solar flare \citep{Chitta2026}. According to fractal-diffusive SOC models, the corresponding differential distributions for flare energy/fluence have a characteristic slope of $\alpha=1.67$ \citep[e.g.][]{AschwandenSchrijver2025}. This value is close to our measured differential slope, $\alpha=1.753$, supporting an SOC-like interpretation of the observed flare statistics over $E_\mathrm{TESS}=10^{29}$--$10^{33}$~erg.

We also test whether the JWST and K2 flare samples require different FFD parameters. We compare a shared single–power-law model to a more flexible model in which JWST and K2 are fit independently. Model selection with the Akaike Information Criterion (AIC) shows that the independently fit model is not favored once the additional degrees of freedom are penalized ($\Delta\mathrm{AIC}=+3.6$). We therefore adopt a single power law for the combined dataset over the energy range shown.

In Figure~\ref{fig:ffd} we also show other FFDs reported in the recent literature. The estimate from TESS TRAPPIST-1 analogs \citep{Seli2021} lies below the TRAPPIST-1 relation at high energies, which may reflect star-to-star dispersion,  sample selection and completeness effects as well as an assumed flare temperature that is too high for ultra-cool dwarfs. JWST-based constraints from \cite{Howard2025}, mapped to the TESS band, span energies of $(2.2$–$8.7)\times10^{30}$~erg and are broadly consistent with our FFD, which covers a much broader range, $10^{29}$–$10^{33}$~erg.

Extrapolating the fitted power law down to $E_\mathrm{TESS}\sim10^{28.5}$\,erg yields a ``microflare-like'' rate that is lower than the independent HST Ly$\alpha$ constraint (green point in Figure ~\ref{fig:ffd}) from \cite{Berardo2025}. It could be attribured to uncteratinties in converting  Ly$\alpha$ variability into a TESS-equivalent energy-rate.

\subsection{The Role of the Flare Spectral Energy Distribution}
A key ingredient for achieving consistency between the two datasets is the assumed flare continuum used to convert flare energies between bandpasses. We adopt white-light flare temperature $T_{\rm flare}=3500$\,K, motivated by JWST observations of flares \citep{Howard2023, Howard2025} and by the expectation that H$_2$ dissociation can act as an effective thermostat in ultra-cool-dwarf atmospheres \citep{ThermostatShapiro}. This cooler flare continuum reduces the \textit{Kepler}-to-TESS energy scaling relative to the canonical 9000--10000\,K assumptions, thereby reducing previously reported offsets between the K2 and JWST cumulative flare rates.  As a systematic check, we also recompute the bandpass conversions over a range of flare temperatures, $T_{\rm flare}=2500-9000$~K (see details in Appendix~\ref{sec:appendix_flare_temperature}). Increasing $T_{\rm flare}$ decreases the inferred TESS-band energy for a given observed flare, shifting the cumulative K2 FFD accordingly and increasing the discrepancy between the cumulative K2 FFD and the JWST FFD.

We also test the sensitivity of the joint JWST+K2 FFD to the assumed flare continuum temperature by repeating the bandpass conversions and model comparison over $T_{\rm flare}=2500$--9000~K. In all cases, model selection favors a shared single power law with $\Delta\mathrm{AIC}\approx+3.8$ at $T_{\rm flare}=2500$~K and $\Delta\mathrm{AIC}\approx+1.6$ at $T_{\rm flare}=9000$~K, indicating that our conclusion that the combined dataset is consistent with a single power law is not driven by the adopted flare temperature.


\subsection{Implications for Flare Energy Budgets and Planetary Irradiation}
Evaluating the fitted cumulative FFD at the threshold $E_\mathrm{TESS}=10^{32}$~erg gives a flare occurrence rate of approximately one event every 25 days. This rate is about an order of magnitude higher than that implied by previous TRAPPIST-1 and TRAPPIST-1-analog FFD estimates at comparable energies \citep{Seli2021,Howard2023}. Thus, the main impact of the shallow single power law is not only to unify the JWST and K2 flare samples, but also to revise upward the expected frequency of energetically important flares affecting the TRAPPIST-1 planets. 
The inferred slope $\alpha<2$ means that the integrated flare energy budget is dominated by the most energetic events.
Consequently, rare flares can drive the strongest transient irradiation. This suggests that flare activity should be treated as a stochastic source of atmospheric forcing, rather than as a small perturbation to the quiescent stellar emission.

\begin{acknowledgments}
VV and NK acknowledge support from the Max Planck Society under the grant ``PLATO Science'' and from the German Aerospace Center under ``PLATO Data Center'' grant   50OO1501. LG acknowledges partial funding from the DLR grant ``PLATO Data Center'' (FKZ 50OO1501 and 50OP1902).
AIS and SS acknowledge support from ERC Synergy Grant REVEAL under the European Union’s Horizon 2020 research and innovation program (grant no. 101118581). AIS and NK acknowledge support from the Volkswagen Foundation (grant 9E126). 
The results reported herein benefited from collaborations and/or information exchange within NASA’s Nexus for Exoplanet System Science (NExSS) research coordination network sponsored by NASA’s Science Mission Directorate.
Based on observations with the NASA/ESA/CSA James Webb Space Telescope obtained from the Mikulski Archive for Space Telescopes (MAST) at the Space Telescope Science Institute, which is operated by the Association of Universities for Research in Astronomy, Incorporated, under NASA contract NAS5-03127.
Support for program number JWST-AR-05370 was provided through a grant from the STScI under NASA contract NAS5-03127.
\end{acknowledgments}

\software{
matplotlib \citep{matplotlib}, 
numpy \citep{numpy}, 
scipy \citep{scipy}
}

\appendix

\section{JWST datasets}  \label{app:data}
The JWST datasets used in this work are summarized in Table~\ref{tab:obs_jwst_summary}.
\begin{deluxetable*}{cccccccc}
\tabletypesize{\small}  
\tablecaption{Summary of JWST Transit Observations of TRAPPIST-1 Planets\label{tab:obs_jwst_summary}.}
\tablehead{
\colhead{Planet} &
\colhead{Program} &
\colhead{PI} &
\colhead{Instrument} &
\colhead{Start Date (UT)} &
\colhead{$\Delta T_\mathrm{obs}$ (hr)} &
\colhead{Flare num.} &
\colhead{Reference}
}
\startdata
b & GO\,2589  & Lim         & NIRISS/SOSS & 2022-07-18 & 4.41   &  & \citet{Lim2023} \\
b & GO\,2589  & Lim         & NIRISS/SOSS & 2022-07-20 & 4.41   & 1 & \citet{Lim2023} \\
c & GO\,2589  & Lim         & NIRISS/SOSS & 2022-10-28 & 4.58   & 2 & \citet{Radica2025} \\
c & GO\,2589  & Lim         & NIRISS/SOSS & 2023-10-31 & 4.58   &  & \citet{Radica2025} \\
f & GTO\,1201 & Lafreni\`ere & NIRISS/SOSS & 2022-10-28 & 3.48 & 3 & Lim et al.\ (in prep.) \\
f & GTO\,1201 & Lafreni\`ere & NIRISS/SOSS & 2023-06-15 & 3.48 & 4 & Lim et al.\ (in prep.) \\
f & GTO\,1201 & Lafreni\`ere & NIRISS/SOSS & 2023-06-24 & 3.48 &  & Lim et al.\ (in prep.) \\
f & GTO\,1201 & Lafreni\`ere & NIRISS/SOSS & 2023-07-03 & 3.48 & 5 & Lim et al.\ (in prep.) \\
f & GTO\,1201 & Lafreni\`ere & NIRISS/SOSS & 2023-07-22 & 3.48 & 6 & Lim et al.\ (in prep.) \\
g & GO\,2589  & Lim        & NIRSpec/PRISM & 2022-07-17 & 4.92   &  7 & Lim et al.\ (in prep.) \\
g & GO\,2589  & Lim        & NIRSpec/PRISM & 2022-12-12 & 4.92   &  8, 9, 10 & Lim et al.\ (in prep.) \\
h & GO\,1981  & Stevenson  & NIRSpec/PRISM & 2023-12-11 & 3.98   & 11 &  Lin et al.\ (in prep.)\\
h & GO\,1981  & Stevenson  & NIRSpec/PRISM & 2023-07-14 & 3.98   & - &   Lin et al.\ (in prep.) \\
e & GTO\,1331  & Lewis     & NIRSpec/PRISM & 2023-06-22 & 3.98   & 12 & \citep{Espinoza2025, Glidden2025}, Lin et al.\ (in prep.) \\
e & GTO\,1331  & Lewis     & NIRSpec/PRISM & 2023-06-28 & 3.98   &  & \citep{Espinoza2025, Glidden2025}, Lin et al.\ (in prep.) \\
e & GTO\,1331  & Lewis     & NIRSpec/PRISM & 2023-07-23 & 3.98   &  & \citep{Espinoza2025, Glidden2025}, Lin et al.\ (in prep.) \\
e & GTO\,1331  & Lewis     & NIRSpec/PRISM & 2023-10-28 & 3.98   & 14 & \citep{Espinoza2025, Glidden2025}, Lin et al.\ (in prep.) \\
c & GO\,2420  & Rathcke   & NIRSpec/PRISM & 2022-07-11 & 4.53   &  & Rathcke et al.\ (in prep.), Lin et al.\ (in prep.) \\
c & GO\,2420  & Rathcke   & NIRSpec/PRISM & 2023-10-29 & 4.51   &  & Rathcke et al.\ (in prep.), Lin et al.\ (in prep.) \\
c & GO\,2420  & Rathcke   & NIRSpec/PRISM & 2023-11-08 & 4.51   & 13 & Rathcke et al.\ (in prep.), Lin et al.\ (in prep.) \\
c & GO\,2420  & Rathcke   & NIRSpec/PRISM & 2024-07-09 & 4.50   &  & \citet{Rathcke2025}, Lin et al.\ (in prep.) \\
\enddata
\end{deluxetable*}

\section{Reduction and Calibration of time-series NIRSpec/PRISM data of TRAPPIST-1} \label{app:b}
The data reduction of NIRSpec/PRISM data of Trappist-1 used in this study was conducted using the package  \texttt{exoTEDRF}, an end-to-end pipeline for JWST exoplanet time series observations \citep{2023Natur.614..670F, 2023MNRAS.524..835R, 2024JOSS....9.6898R, 2025ApJ...985L..10A}.

The pipeline processes the raw detector data by identifying saturated pixels and flagging the affected columns for subsequent reduction steps, and by applying bias and non-linearity corrections. Time-correlated 1/f noise and the additive background were corrected at the group level using column-wise statistics of non-illuminated pixels. Most NIRSpec/PRISM observations used the 32-pixel-wide SUB512 subarray, except for the first three visits of GTO 2420 (PI: Rathcke), which employed the 16-pixel-wide SUB512S subarray on 2022 July 11, 2023 October 29, and 2023 December 8. For these visits, background estimation excluded a $\pm10$-pixel width region around the spectral trace, while a $\pm14$-pixel width exclusion was used for all other observations. 

The pipeline then identifies and corrects jumps (e.g., cosmic ray hits) using a time-domain sigma-clipping algorithm with a 7$\sigma$ rejection threshold. 
Ramp fitting is subsequently performed at the integration level using the corrected group-level reads. Finally, after bad pixel identification and correction, the reduced data were then used to extract 1-dimensional spectra with a 6-pixel width box and apply wavelength calibration.

The flux density units of NIRSpec/PRISM Trappist-1 spectra were absolutely calibrated using a flux-calibration sensitivity function derived from NIRSpec/PRISM observations of the A-type star 2MASS~J17571324+6703409 obtained by the CAL/CROSS 1536 program (PI: Karl Gordon; \citealt{2019jwst.prop.1536G}). 
The 32-pixel-wide SUB512 subarray data of the A-type star were reduced using the same pipeline and procedures described above. 
The flux-calibration sensitivity curve was constructed by comparing the mean of the extracted count-rate spectra with the absolute flux spectrum of 2MASS~J17571324+6703409 from the \texttt{CALSPEC} library \citep{2014PASP..126..711B, 2020AJ....160...21B, 2022AJ....164...10B, 2025AJ....169...40B}.

\section{Detection efficiency in the JWST observations}\label{sec:appendix_detection_eff_jwst}

We quantify the flare detection efficiency (completeness) using an injection--recovery experiment performed on the  H$\alpha$ light curves for each JWST visit. We treat the JWST/NIRISS and JWST/NIRSpec datasets separately and report completeness as a function of the inferred TESS-band flare energy, $E_{\rm TESS}$.

\subsection{Injection templates and scaling}
\begin{figure}[htbp]
  \centering
  \includegraphics[width=\linewidth]{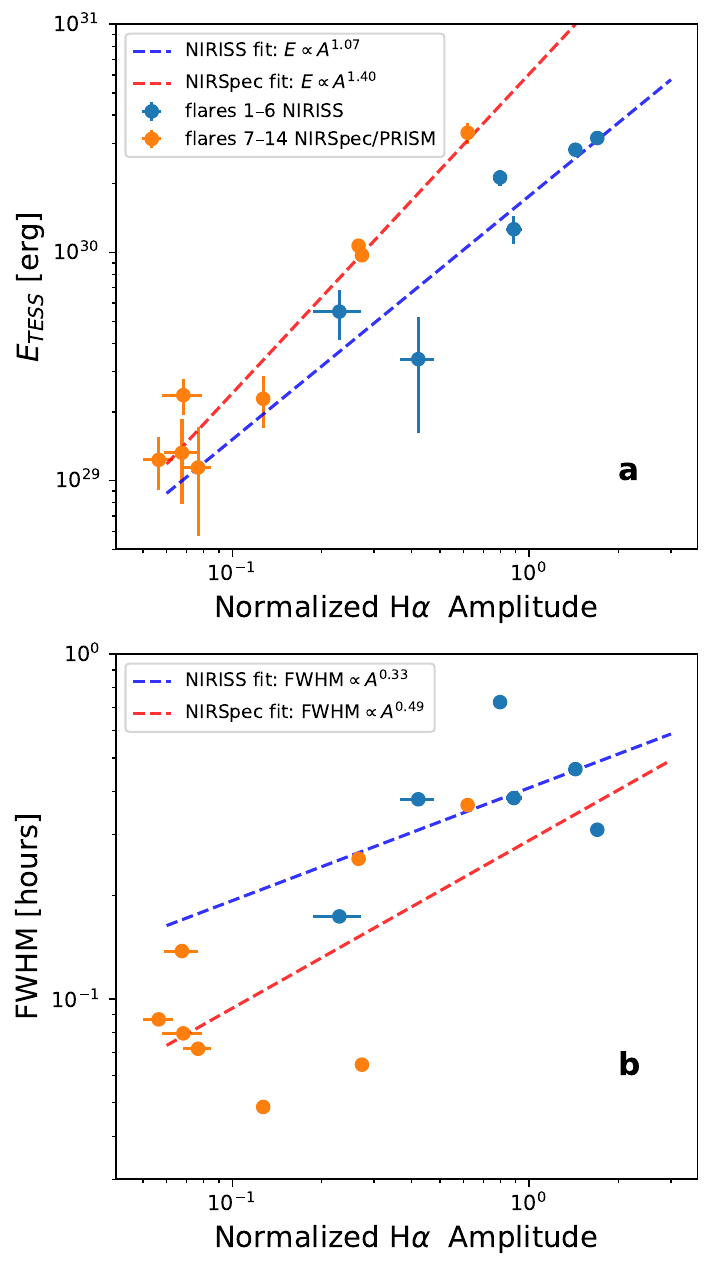}
  \caption{\textbf{Amplitude--Energy and Amplitude--Duration Relations for JWST flares.}
  \textbf{(a)} Relationship between normalized H$\alpha$ flare amplitude and inferred $E_{\rm TESS}$ and \textbf{(b)} between amplitude and flare duration (FWHM) for the NIRISS and NIRSpec/PRISM flare samples. Dashed lines show linear fits in log--log space. The energy--amplitude mapping is steep and relatively tight, whereas the duration--amplitude trend is weaker and more scattered; we therefore scale injected flares in amplitude only for the injection--recovery test.
}
  \label{fig:amplitude_tess_versus_halpha_amplitude}
\end{figure}
We adopt a data-driven approach and use representative flares observed in our H$\alpha$ datasets as empirical templates. Specifically, we take Flare~2 ($E_{\rm TESS, ref}=2.8\times10^{30}\,$erg) as the reference H$\alpha$ template for the NIRISS dataset and Flare~8 ($E_{\rm TESS, ref}=1.1\times10^{30}\,$erg) for the NIRSpec dataset.

To probe lower energies we scale the reference H$\alpha$ flare template by a positive amplitude factor from $a=0.05$ to $1$ while keeping its time profile fixed. This injection scheme is designed to isolate completeness as a function of flare amplitude for representative, empirically observed flare morphologies, without imposing an assumed flare energy--duration scaling at these wavelengths and cadences. The choice of a fixed time profile is motivated by the limited number of detected JWST flares and by the weak and scattered dependence of flare duration on amplitude in our sample (see Fig.~\ref{fig:amplitude_tess_versus_halpha_amplitude}), which does not support adopting a unique amplitude--duration relation for injections. Moreover, the physically representative range of flare widths at fixed H$\alpha$ amplitude is not well constrained at these wavelengths and cadences. Because the injected excess flux is multiplied by $a$ at all times, the time-integrated flare signal (and hence the inferred $E_{\rm TESS}$ for our adopted energy estimator) scales linearly,
\begin{equation}
E_{\rm TESS}(a)=a\,E_{\rm TESS,ref}.
\end{equation}

For each trial, we inject one scaled flare template at a random time (uniformly distributed over the observing window) and run the flare detection pipeline (described below). The pipeline identifies all candidate flare peaks that exceed the adopted detection threshold in the matched-filter time series. In the presence of noise and residual systematics it can return multiple candidates even when only a single flare is injected. Our completeness statistic is therefore defined in the standard way for injection--recovery tests: an injection is considered recovered if the pipeline returns at least one detection consistent with the injected event. Additional detections are not counted as additional recoveries. To prevent real astrophysical flares in the original data from biasing the recovery statistics, we apply a differential matching criterion:
\begin{enumerate}
\item First, we remove any detected candidates that fall within a veto window around the peak times of known real flares in that light curve. This step avoids counting pre-existing events as "recoveries" of the injected flare. Equivalently, one could restrict injections to non-veto intervals. Our implementation retains a uniform injection distribution in time and applies the veto at the matching stage. 
\item An injection is counted as recovered if at least one of the remaining detections falls within a matching tolerance $\Delta t_{\rm rec}$ of the injected peak time. Multiple detections within the matching window are counted once.
\end{enumerate}

We adopt $\Delta t_{\rm rec}=4\,dt$ so that a recovered flare is allowed to land within a few cadence steps of the injected peak time, where $dt$ is the cadence of the H$\alpha$ light curve. This accommodates the typical peak-timing uncertainty introduced by noise and finite sampling, while keeping the matching window small enough to avoid counting chance coincidences as recoveries.

\subsection{Completeness curves}

For the H$\alpha$ equivalent of a given injected energy $E_{\rm TESS}$, we perform $N=1000$ trials and compute the recovery fraction
\begin{equation}
p(E_{\rm TESS}) \;=\; \frac{K}{N},
\end{equation}
where $K$ is the number of recovered injections. We repeat this procedure for all visits in a given instrument dataset and report the dataset-averaged completeness curve (i.e., the mean $p$ at fixed $E_{\rm TESS}$ across visits).

\begin{figure}[htbp]
  \centering
  \includegraphics[width=1.0\linewidth]{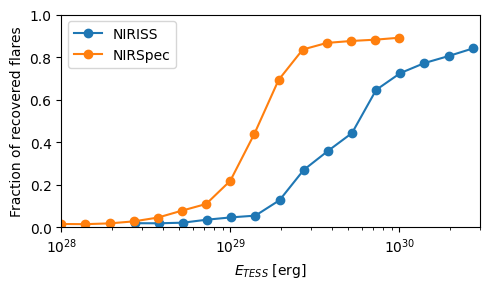}
  \caption{\textbf{Completeness curves for the JWST datasets.}
  Dataset-averaged recovery fraction $p(E_{\rm TESS})$ for the JWST/NIRISS (blue) and JWST/NIRSpec (orange) injection--recovery experiments as a function of injected flare energy.}
  \label{fig:fraction_of_detected_flares}
\end{figure}

Figure~\ref{fig:fraction_of_detected_flares} shows the resulting dataset-averaged recovery fractions for NIRISS and NIRSpec. Because we allow injections to occur at arbitrary times, including intervals that contain real flares, the completeness does not reach unity. In particular, when an injected flare overlaps a real flare, the differential matching procedure can remove detections associated with the real event, making the injected flare effectively unrecoverable; this produces a saturation level of $\sim 0.8$--$0.9$ rather than 1.0. The curves also show that the NIRSpec dataset is sensitive to lower-energy events than NIRISS, with ${\sim}50\%$ recovery at $E_{\rm TESS}\sim6\times10^{29}\,$erg for NIRISS and $E_{\rm TESS}\sim1.5\times10^{29}\,$erg for NIRSpec. We use these completeness curves to correct the observed FFD.

\subsection{Matched-filter detection statistic}\label{sec:mf}
Our flare search uses a matched-filter approach: we slide a parametric flare template through each normalized H$\alpha$ light curve and quantify, at every trial time, how significantly the data prefer a flare-shaped brightening over a flat local baseline. We do this as a function of FWHM and look for peaks in the two-dimensional cross-correlation (as a function of trial time and FWHM). We model the flare signal as
\begin{equation}
m(t;A,t_0,\mathrm{FWHM}) = A\,h(t;t_0,\mathrm{FWHM}),
\end{equation}
where $h(t;t_0,\mathrm{FWHM})$ is a unit-amplitude flare template with peak time $t_0$ and timescale parameterized by its FWHM, and $A$ is the fitted amplitude.

We start from the analytic template of \citet{Mendoza2022}, which was originally calibrated on optical (Kepler/TESS) white-light flares. To better approximate the morphology of the H$\alpha$ flares in our JWST time series, we apply a simple asymmetric time-warp to the template. We extend the rise-time axis by a factor of 4 and compress the decay-time axis by a factor of 10 relative to the original Mendoza template. This warping preserves the template’s unimodal shape while allowing a less impulsive rise and a more rapidly decaying tail, consistent with the H$\alpha$ flare profiles observed in our data. This empirical warping is not intended as a physical model of H$\alpha$ radiative transfer: it is a practical choice that improves peak localization and detection sensitivity for our cadence and S/N.
The matched-filter detection statistic is then computed using this warped unit-amplitude template evaluated on the observation timestamps. With these choices, the detection pipeline recovers the 14 flares identified by visual inspection at the correct times, and we adopt the same template for the injection--recovery completeness analysis.

For each trial pair $(t_0,\mathrm{FWHM})$ we estimate a local baseline as the inverse-variance weighted mean flux in a symmetric window of half-width $5\times\mathrm{FWHM}$ around $t_0$. We subtract this baseline to form residuals, and we also subtract the template's weighted mean within the same window to avoid sensitivity to constant offsets. The flare amplitude $A$ is then obtained by weighted least squares (weights $w_i=1/\sigma_i^2$), and the detection statistic is taken as the corresponding matched-filter signal-to-noise ratio (fitted amplitude divided by its predicted 1$\sigma$ error from the weighted least-squares fit). We require flare-like brightening by enforcing $A>0$.

We evaluate the matched filter on a two-dimensional grid:
(i) a logarithmically spaced $\mathrm{FWHM}$ grid with 50 values spanning from $0.01$~h (in the same time units as the light curve; chosen to be slightly above the cadence) up to half the light-curve duration; near the edges the $\pm 5\times\mathrm{FWHM}$ window is truncated to the available data and trials with fewer than $N_{\rm min}=5$ points are skipped;
(ii) trial peak times are scanned on a uniform grid spanning the full light-curve duration, with $3\times$ oversampling relative to the number of data points in the light curve), to allow flare peaks $t_0$ to occur between sampled timestamps. 

\begin{figure*}[htbp]
  \centering
  \includegraphics[width=\linewidth]{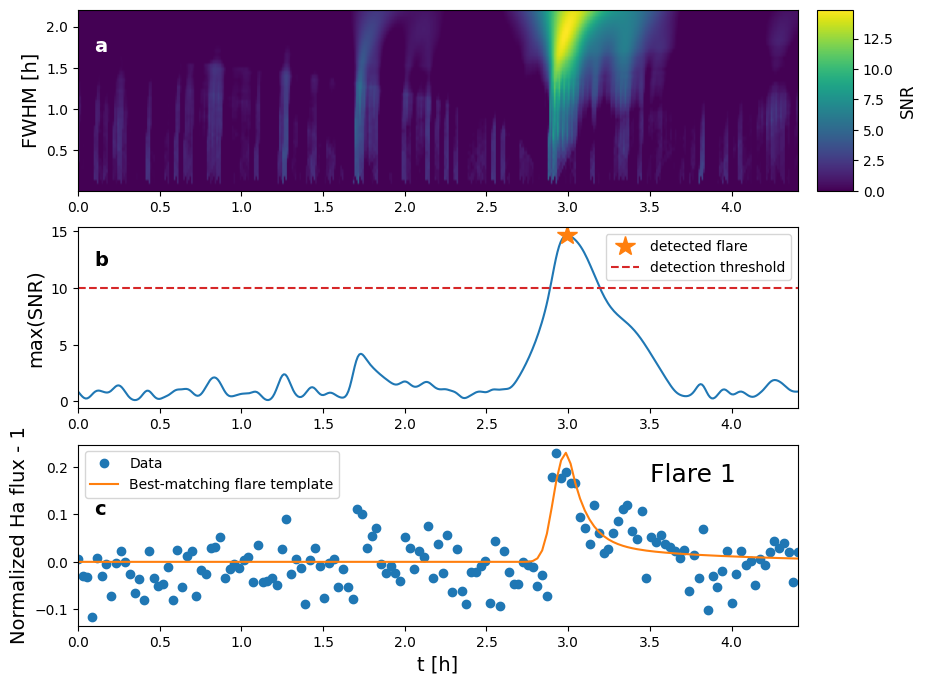}
  \caption{\textbf{Matched-filter flare detection illustrated for Flare~1.}
  (a) Two-dimensional matched-filter S/N map as a function of trial peak time $t_0$ and template $\mathrm{FWHM}$.
  (b) One-dimensional detection statistic obtained by maximizing the S/N over $\mathrm{FWHM}$ at each $t_0$ and then smoothing in time; the horizontal dashed line marks the adopted detection threshold (S/N $=10$), and the star marks the detected peak.
  (c) H$\alpha$ light curve around the detected event (points) together with the best-matching flare template (curve) corresponding to the peak in panel~(b). This event occurs during a planetary transit. However, no explicit transit subtraction is applied in H$\alpha$, since stellar variability dominates at these wavelengths and cadences.
  }
  \label{fig:flare_cc_detection_example}
\end{figure*}

At each trial time $t_0$ we retain the most significant timescale by taking the maximum matched-filter S/N over the $\mathrm{FWHM}$ dimension, yielding a one-dimensional detection statistic as a function of time. To suppress isolated noise excursions, we smooth this 1D S/N series with a Gaussian kernel whose width is set to the light-curve cadence. Candidate flares are then identified as local maxima of the smoothed S/N series that (i) exceed a fixed detection threshold (S/N $>10$), (ii) have a minimum peak prominence of 1.0, and (iii) are separated by at least a small fraction of a characteristic flare timescale (set to $0.1$ times the median value of the $\mathrm{FWHM}$ grid) to avoid multiple triggers on the same event. For each detection we report the trial peak time $t_0$ and the $\mathrm{FWHM}$ at which the original 2D S/N map attains its maximum, adopting these as the flare time and a proxy for its duration. Figure~\ref{fig:flare_cc_detection_example} summarizes the main steps of the flare detection procedure for Flare~1.

\section{Choosing the lower threshold $E_{\min}$}\label{app:emin}
Our goal in selecting $E_{\min}$ is to isolate the high-energy regime of the cumulative FFD from the K2 data that is approximately linear in log-log space (i.e., power-law-like), and to exclude lower-energy flares where incompleteness become apparent. We therefore treat a power-law fit as a convenient local parametrization of the tail, rather than asserting that the full distribution is generated by an exact power law.

\begin{figure*}[htbp]
  \centering
  \includegraphics[width=\linewidth]{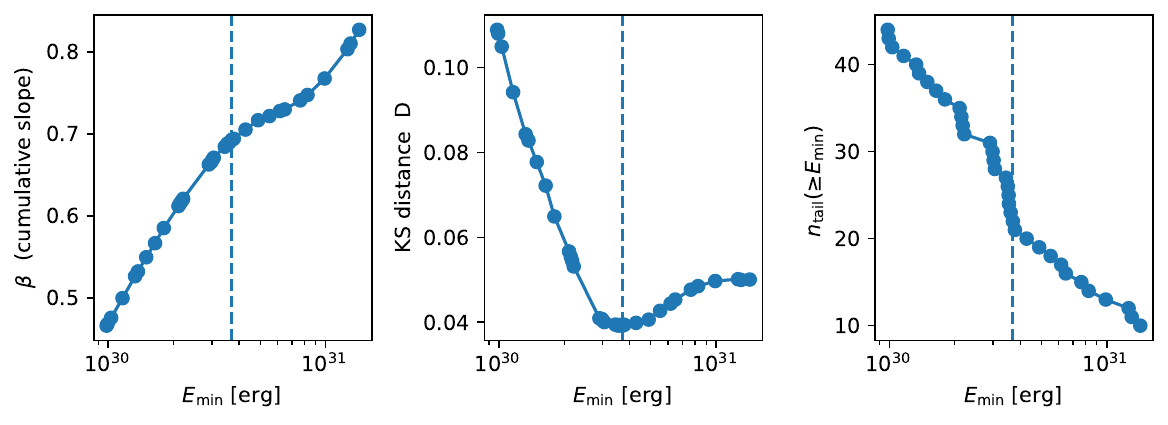}
  \caption{\textbf{Selection of the lower energy threshold $E_{\min}$ for the K2 high-energy tail.}
  Left: fitted tail slope $\beta$ as a function of $E_{\min}$.
  Middle: KS distance between the empirical tail ($E_\mathrm{TESS}\ge E_{\min}$) and the corresponding best-fit power-law tail, used here as a diagnostic of how well the high-energy tail is approximated by a linear relation in log-log space.
  Right: number of flares retained in the tail, $n_{\rm tail}$.
  The adopted $E_{\min}=3.6\times 10^{30}$~erg (vertical dashed line) is chosen to minimize  the KS distance while yielding a stable slope and retaining sufficient statistics.}
  \label{fig:emin_kepler}
\end{figure*}

To choose $E_{\min}$, we scan candidate thresholds and, for each value, fit a straight line to the cumulative FFD in log-log space using only events with $E_\mathrm{TESS}\ge E_{\min}$. We then monitor three diagnostics as functions of $E_{\min}$ (Fig.~\ref{fig:emin_kepler}): (i) the fitted tail slope $\beta$, (ii) the Kolmogorov-Smirnov (KS) distance between the empirical tail and the corresponding best-fit power law  (used here as a goodness-of-linearity diagnostic), and (iii) the number of flares retained in the high-energy tail, $n_{\rm tail}(\geq E_{\min})$. We adopt $E_{\min}$ at the point where the KS distance is minimized  while the inferred slope is stable against small changes in $E_{\min}$ and the tail still contains a sufficient number of events for a meaningful fit.

With this choice, the selected subset $E_\mathrm{TESS}\ge E_{\min}=3.6\times 10^{30}$~erg represents the approximately power-law part of the FFD, without requiring the power-law model to be exact outside this high-energy regime.

\section{Impact of the flare temperature on the flare frequency distribution derived from white-light observations }\label{sec:appendix_flare_temperature}
\begin{figure*}[htbp]
  \centering
  \includegraphics[width=0.9\linewidth]{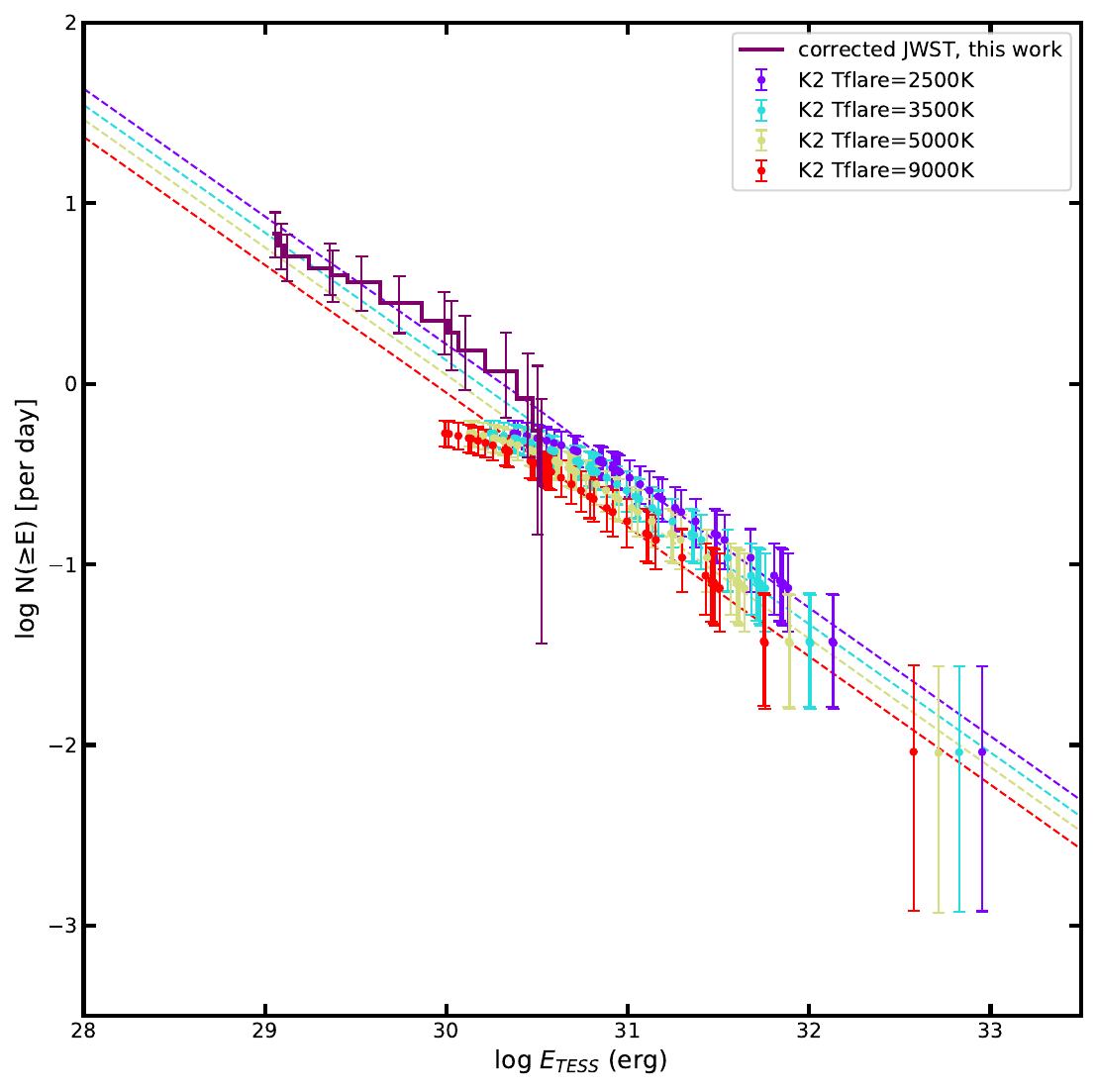}
  \caption{\textbf{Impact of the assumed flare temperature on the cumulative flare frequency distribution}  Colored markers show the K2 cumulative FFD recomputed for different assumed flare temperatures, and dashed lines show power-law fits to the high-energy regime.
For a fixed observed flux enhancement, higher $T_{\rm flare}$ implies a smaller inferred TESS-band flare energy, shifting the K2-based FFD to lower $E_{\rm TESS}$. In particular, adopting a solar-like white-light temperature ($T_{\rm flare}\sim 9000$~K) shifts the K2 FFD sufficiently that its extrapolation becomes inconsistent with the FFD from JWST observations.
}
  \label{fig:ffd-temperature}
\end{figure*}

The goal of this section is to quantify how the inferred flare frequency distribution (FFD), particularly for the K2 sample, depends on the assumed flare effective temperature. We consider four temperatures: $T_{\rm flare}=2600$~K (close to the stellar effective temperature), 3500~K (the fiducial value adopted in the main analysis following \citealt{ThermostatShapiro}), 5000~K (an intermediate value), and 9000~K (representative of typical white-light flares on Sun-like stars). For each temperature we recomputed the FFD and fitted a power law to the high-energy portion of the cumulative distribution (Fig.~\ref{fig:ffd-temperature}). We then extrapolated the best-fit relation to the energies measured for the JWST events.

Increasing $T_{\rm flare}$ decreases the inferred TESS-band energy associated with a given observed flare, shifting the K2-based cumulative FFD toward lower energies and thereby modifying its apparent alignment with the JWST constraints. Across the explored range $T_{\rm flare}=2500$--9000~K, the best-fit cumulative slope steepens modestly from $\beta=0.714^{+0.061}_{-0.052}$ to $\beta=0.841^{+0.090}_{-0.061}$.

Our choice of $T_{\rm flare}=3500$~K is motivated by independent JWST constraints on TRAPPIST-1 flare continua at red optical/near-IR wavelengths \citep[e.g.,][]{Howard2023,Howard2025}, and by theoretical expectations that H$_2$ dissociation can regulate flare temperatures in ultra-cool-dwarf atmospheres \citep{ThermostatShapiro}. We note that adopting cooler continua ($T_{\rm flare}\sim2500$--3500~K) yields the closest agreement between the K2 and JWST cumulative FFDs in Fig.~\ref{fig:ffd-temperature}, whereas hotter assumptions ($T_{\rm flare}\gtrsim5000$~K) increase the discrepancy.

\bibliography{bibliography}
\bibliographystyle{aasjournalv7}

\end{document}